\documentstyle[aas2pp4]{article}

\begin{document}

\title{Energy Spectra and High Frequency Oscillations in 4U~0614+091}

\authoremail{eric@astro.columbia.edu}

\author{E.C. Ford\altaffilmark{1}, P. Kaaret\altaffilmark{1},
K.Chen\altaffilmark{1}, M.Tavani\altaffilmark{1},  D.
Barret\altaffilmark{2,3}, P. Bloser\altaffilmark{3},  J.
Grindlay\altaffilmark{3}, B.A. Harmon\altaffilmark{4},  W.S.
Paciesas\altaffilmark{5,4}, and S.N. Zhang\altaffilmark{6,4} }

\altaffiltext{1}{Department of Physics and Columbia Astrophysics
Lab, Columbia University, 538 W. 120th Street, New York, NY10027}  
\altaffiltext{2}{Centre d'Etude Spatiale des Rayonnements (CESR), 
9 Avenue du Colonel Roche BP4346, 31028 Toulouse Cedex 4, FRANCE} 
\altaffiltext{3}{Harvard Smithsonian Center for Astrophysics, 
60 Garden Street, Cambridge, MA 02138}
\altaffiltext{4}{NASA/Marshall Space Flight Center, ES 84,
Huntsville, AL 35812} 
\altaffiltext{5}{University of Alabama in Huntsville, Department of 
Physics, Huntsville, AL 35899}
\altaffiltext{6}{Universities Space Research Association/MSFC, ES 84, 
Huntsville, AL 35812}

\begin{abstract}

We investigate the behavior of the high frequency quasi-periodic
oscillations (QPOs) in 4U~0614+091, combining timing and
spectral analysis of  RXTE observations. The energy spectra
of the source can be described by a power law ($\alpha \sim 2.8$)
and a blackbody ($kT \sim 1.5$), with the blackbody accounting for 10--20\% 
of the total energy flux. We find a robust correlation of the frequency,
$\nu$, of the higher frequency QPO near 1 kHz with the flux of the 
blackbody, $F_{BB}$. The slope of this correlation, 
$d\log{\nu}/d\log{F_{BB}}$, is 0.27 to 0.37.
The source follows the same relation even in observations separated by 
several months. The QPO frequency does not have a similarly unique correlation 
with the total flux or the flux of the power law component.  The RMS 
fraction of the higher frequency QPO rises with energy 
from $6.8\pm1.5$\% (3--5 keV) to $21.3\pm4.0$\% (10--12 keV). For the lower 
frequency QPO, however, it is consistent with a constant value of
$5.4\pm0.9$\%. The results 
may be interpreted in terms of a beat frequency model for the production 
of the high frequency QPOs.

\end{abstract}


\section{Introduction}

Recent observations with the Rossi X-ray Timing Explorer (RXTE)
have revealed quasi-periodic oscillations (QPOs) at high
frequencies in numerous low mass X-ray binaries (see van der Klis
1997a, Ford et al. 1997b). These oscillations
are likely produced within a few stellar radii of the surface of
the neutron star; a typical frequency of 1000 Hz corresponds to
an orbital  radius of 15 km for matter in Keplerian motion
around a $1.4 M_{\odot}$  neutron star. The behavior of these
QPOs is yielding information on fundamental properties of the
neutron stars: measurements of their spin periods (Strohmayer 
et al. 1996, Ford et al. 1997a) and masses (Kaaret, Ford, \& Chen
1997; Zhang, Strohmayer, \& Swank 1997b) and constraints on
their radii (Miller, Lamb \& Psaltis 1997). High frequency QPOs
may also serve as  unique probes of strong field general
relativity (Kaaret, Ford,  \& Chen 1997).

One area of analysis, which has not been exploited to date, is
the study of the energy spectra. Combining the spectral behavior
with the QPO properties may prove particularly fruitful. Here
we present such an analysis for 4U~0614+091, one of the X-ray 
bursters with strong QPOs near 1000 Hz (Ford et al. 1997a, 
Mendez et al. 1997). Section 2 is a description of the observations 
and the results of the energy spectral analysis. In Section 3 we 
correlate the spectral fits with the QPO behavior. In Section 4 we 
interpret these results in the context of a magnetospheric beat 
frequency model.

\section{Observations and Spectral Analysis}

RXTE observations of 4U~0614+091 were conducted on 22 and 24
April and 7 and 8 August 1996 for a total usable time of 79
ksec. Details of the observations and analysis of the high frequency
QPOs are presented in Ford et al. (1997a). RXTE has two coaligned
pointed  instruments. The proportional counter array (PCA) consists of
five separate proportional counter units (PCUs) with a large
effective area from about 2 to 20 keV peaking at about 7000 cm$^{2}$. 
Each PCU consists of three Xe layers stacked  behind a propane 
veto layer. The second pointed instrument, HEXTE, is a NaI array 
and extends energy coverage from 15 to 100 keV.

The accuracy of the spectral fits is currently dominated by 
systematic errors. We have found that the largest of these are the
uncertainties in the PCU detector response matrices. Using various
recent matrices, for example, we find that the total flux of the fitted
spectra can change by 30\%. Here we use the latest available 
matrices (Jahoda et al. 1997), which have been tested using the Crab 
continuum and the Fe line from Cas A. 

We fit all five PCUs simultaneously and allow a varying
normalization factor between detectors. We also use only the 
topmost Xe layer in which the source rates are highest, though the
fits do not depend sensitively on which layers are used. We use the
background estimation in the PCU based on particle activation
and cosmic background (Stark et al. 1997). The background
subtraction for the HEXTE detectors is based on the on-source, off-source 
rocking of the detectors.


A typical spectral fit from 4U~0614+091 is summarized in Table~1.
We consider here a two-component continuum model consisting of a
power law (PL) and blackbody (BB).
This is an appropriate description of the continuum; such
models have been used for X-ray binaries (White, Stella and Parmar 
1988, Christian \& Swank 1997), and previous observations of
4U~0614+091 with EXOSAT (Singh and Apparao 1994,  Barret and Grindlay 
1994) and Einstein SSS (Christian, White and Swank 1994) have 
suggested a BB plus PL model with similar values of PL index,
BB temperature, and fluxes.

The BB plus PL description can be compared to other available
continuum models, though a strict comparison using $\chi^2$ statistics
is not possible at this time since the fits are dominated by details of the 
detector response matrices. It is possible to artificially reduce the
$\chi^2$ values by adding an extra error (e.g. 1\%) to each energy
bin. We have not done this. Single component models, such as a BB or PL 
(Table~1), are less appropriate than the BB plus PL, as there are 
larger residuals with peculiar shapes and the $\chi^2$ are larger.
Other models are also less favored: a modified disk blackbody (Stella
and Rosner  1984) and the disk model of Mitsuda et al. (1984) produce
large residuals at low and high energies in the PCA, and have
$\chi^2/\nu$ values of 3065/259 and 12345/258 respectively. A Compton
scattering model (Sunyaev and Titarchuk 1980) provides a somewhat
better fit, but still yields $\chi^2/\nu = 2758/258$.

Over our observations (more than 25 orbits), 4U~0614+091 shows
significant variability. The flux  of the PL component covers
the range $17.1-24.0\times10^{-10}$ erg cm$^{-2}$ s$^{-1}$ (1--20
keV), with typical errors of 5\% while the BB flux
varies from $1.5-6.4\times10^{-10}$ erg cm$^{-2}$ s$^{-1}$
(errors of $\sim$3\%), which is 8 to 21 \% of the total flux.
As the BB flux increases, the photon index of the power law softens 
from 2.62 to 3.17 (typical error 0.02). 
This trend is opposite to the one expected if the two components 
are actually just a non-physical description of the continuum shape.
For an accidental coupling, we would expect the blackbody flux to 
increase as the powerlaw hardens to take over the missing flux 
at low energy.

We are excluding here the data during 22 April and 8 August
observations. In these intervals the blackbody components have low
fluxes (less than about $1\times10^{-10}$ erg cm$^{-2}$ s$^{-1}$) and
are cool ($kT < 0.7$ keV) and not well determined in the PCUs, as
we impose a lower energy bound of 2 keV due to the poor response of
the PCUs at low energy. Spectral parameters for these intervals 
have proven unreliable.

Adding interstellar absorption can change the flux of the
blackbody by as much as 5\%. The value of equivalent
hydrogen column density, $n_H$, is not well determined due to the
cutoff at 2 keV. We therefore fix $n_H$ at 
$0.18\times10^{-22}$ cm$^{-2}$, a value  consistent with 
the previous observations (Christian, White \& Swank 1994). In
establishing errors  we have allowed $n_H$ to vary between 
$0.13-0.23\times10^{-22}$ cm$^{-2}$, a range that encompasses
the values from previous observations.

The spectral fits in all observations are improved with the 
addition of a narrow line at an energy of about 7 keV. 
A feature at the same energy has also been identified in the 
EXOSAT data (Singh and Apparao 1994). We have not included this
line in the fits quoted here, but the results below do not
depend sensitively on whether this line is included.

The fits are also insensitive to the inclusion of HEXTE data.
We quote results here including the HEXTE data, but removing HEXTE
changes the fluxes and fit parameters by less than 1\%.

\section{Correlation of Spectral and Timing Variations}

The two high frequency QPOs in 4U~0614+091 are simultaneously
present in most observations and are separated in frequency  by
323 Hz (Ford et al. 1997a). Here we focus mainly on the higher
frequency QPO. Figure~1 shows that the higher QPO frequency,
$\nu$,  correlates very well with the flux of the blackbody
spectral component, $F_{BB}$. 

The exact details of this correlation depend on the spectral fitting.
In particular changing the response matrices changes the flux estimates.
We note, however, that the correlation is apparent in all trials
using different response matrices and models. We parameterize the 
correlation in terms of the slope, $\alpha=d\log{\nu}/d\log{F_{BB}}$. 
The fits in Figure~1 yield $\alpha=0.29 \pm 0.01$. In our various spectral
fitting trials, $\alpha$ is always in the range 0.27 to 0.37. The slope 
depends somewhat on the energy integration range used in calculating 
the BB fluxes. We have used a range 1--20 keV. Though this range extends 
to somewhat beyond the PCA window, using such a wide band makes $\alpha$ 
insensitive to the particular choice of energy cutoffs.

The higher QPO frequency clearly correlates well with BB flux. This 
is in contrast to the variations of QPO frequency with respect to count 
rate, $R$, where the August and April 1996 observations occupy
two separate branches in the $R-\nu$ plane (Ford et al. 1997a).
Figure~2 shows there is no unique correlation of frequency with 
total flux. Similarly the flux of the PL does not uniquely determine
the QPO frequency. Thus, only the BB flux can be regarded as a 
good indicator of QPO frequency.

Figure~3 shows the energy resolved fractional RMS amplitude for
both QPO features. The data are taken from an interval with one
of the  stronger QPO detections. The fractional amplitude of
the higher frequency QPO clearly rises up to about 12 keV, which
indicates that the spectrum associated with the oscillation is
harder than the time averaged spectrum. The fractional amplitude of
the lower frequency QPO is approximately constant with energy. 
Figure~3 suggests that the two QPOs have different trends
of RMS fraction with respect to energy. If the shapes are
in fact distinguishable, then this different energy dependence may 
provide an independent diagnostic for identifying which of the two 
QPOs is present when only one QPO is observed from a source.

\section{Discussion}

The results of Section~3 show that the QPO frequency is uniquely
determined by the flux of the blackbody component of the
spectrum  and not the total flux or the flux of the power law
component. As a working hypothesis we consider a beat-frequency
interpretation in which the higher frequency QPO is determined by 
the orbital Keplerian frequency of matter in the inner disk.

In the magnetospheric model (Alpar \& Shaham 1985) the disk is
interrupted by the pressure of the magnetosphere (Ghosh \&
Lamb 1979), and the Keplerian frequency and luminosity scale as
$\nu_K \propto L^{\alpha}$. If the luminosity is released at the neutron
star surface, then the value of $\alpha$ is 3/7.  If the energy 
is liberated mainly at the magnetospheric boundary, then 
$\alpha=1/3$ (Alpar and Shaham 1985, Lamb et al. 1985). These
numbers are derived assuming a pure magnetic  dipole field and a simple
form for the luminosity. Our determination of $\alpha=0.27$ to 0.37 
is remarkably close to these simple estimates. The measurement, however,
is not sufficiently well constrained to distinguish between the two
predictions for $\alpha$.

The correlation of QPO frequency with blackbody flux indicates that 
the radius of the inner disk is related to the blackbody flux. Both
the blackbody flux and the inner disk radius, apparently, are determined 
by the mass accretion rate through the disk. This picture is consistent
with the behavior of the inferred size of the emission region. The size 
grows as the flux and frequency increase. This is as expected since the
x-ray emitting area of the disk increases in size as the inner disk
radius shrinks.

The emission in the QPO itself, however, is probably 
not from the blackbody. This is because the RMS fraction increases 
at higher energies, which means that the QPO emission is harder than 
the average spectrum. Given the correlation of QPO frequency with 
blackbody flux, this might seem paradoxical. We note, however, that 
the blackbody flux may track the QPO frequency (by a common dependence 
on mass accretion rate for instance) while not actually producing the 
photons in the modulation.

The QPO photons are more likely associated with the emission in the power 
law component. The power law component, which carries most of the energy, 
could be produced in the inner-most regions as matter falls from the 
inner disk edge to the neutron star surface (Kluzniak \& Wagoner 1985). 
Alternatively, non-thermal magnetic effects may produce the power 
law (Tavani \& Liang 1996).

The evidence for a beat frequency mechanism for the kiloHertz QPOs
is compelling.
In many atoll sources so far observed, high frequency QPOs are 
detected simultaneously at two frequencies. The difference in
the frequencies of the two QPOs remains constant  over a span of
months,  as confirmed in several sources  (Ford et al. 1997a,
Strohmayer et al.  1996). This is exactly as predicted by the
beat frequency model; the higher frequency QPO is from Keplerian
motion, and the lower frequency QPO is a modulated interference
with the spin of the neutron star. The frequency difference between
these two signals constant and equal to the neutron star spin
frequency. Remarkably, a third signal has been detected during X-ray 
bursts in a few sources at a frequency equal to the difference 
frequency or a multiple of the difference (Strohmayer et al. 1996,
Smith, Morgan \& Bradt 1997, Wijnands \& van der Klis 1997b). This 
suggests that the difference 
frequency is indeed the spin frequency of the neutron star.
In 4U~1636-536 the frequency of  oscillations discovered during
a burst is marginally consistent with  twice the difference
frequency (Zhang et al. 1997a, Wijnands et al.  1997a). A QPO at
the difference frequency has been reported in the quiescent
emission of 4U~0614+091 (Ford et al. 1997a). The behavior of the
high frequency QPOs in atoll sources is consistent with a beat
frequency description.

A glaring exception to this behavior is the Z-source Sco X-1 (van
der Klis et. al 1997b). In Sco X-1, the QPOs near 1~kHz are
broadly similar to those in the atoll sources discussed above,
but here the difference of the QPOs is not constant but decreases with 
increasing luminosity (van der Klis et al. 1997b). This would
appear to be a case against  the simple beat frequency model. 
Perhaps the effects of radiation are more important in Sco X-1 
where the luminosity is about 100 times larger than in 4U~0614+091.

Recently QPOs near 1000 Hz were discovered in GX~5-1 (van der
Klis et al. 1996) and GX~17+2 (van der Klis et al. 1997c), 
archetypal strong Z-sources which are known to have  strong low 
frequency horizontal branch QPOs. These results put the original
magnetospheric beat frequency model in an awkward position. The
original model (Alpar \& Shaham 1985, Lamb et al. 1985) was
adopted to explain the 15 to 60 Hz horizontal  branch
oscillations in Z-sources. These bright sources were
observed with only  single QPOs, presumed to be the beat frequency.
It is still puzzling that  the other two frequencies implied by
the model have never been observed. Both the kHz oscillations and the 
horizontal branch oscillations cannot arise at a single inner disk. 
The applicability of the models to horizontal branch oscillations
needs further examination. Certainly, the conclusions and constraints 
on the beat frequency model from the horizontal branch QPOs should 
not be used against its application to high frequency QPO observations.

~~

We would like to thank the staff of the RXTE Guest Observer Facility,
in particular Keith Jahoda for discussion and support on the PCA response. 
This work was supported by a NASA RXTE grant and a NASA Graduate Student 
Researchers Program Award.


\begin{deluxetable}{llllllll}

\tablewidth{45pc}  \tablecaption{Spectral Fits}
 
\tablehead{ \colhead{MODEL} & \colhead{$\alpha$} &
\colhead{Norm} & \colhead{$F_{PL}$\tablenotemark{a}} &
\colhead{$kT_{BB}$} & \colhead{Norm} &
\colhead{$F_{BB}$\tablenotemark{a}} &
\colhead{$\chi^{2}$/dof} \nl 
 & & & & (keV) & & &  }
 
\startdata
BB              & ...   & ...   & ... & 1.244 & $1.29\times10^{-2}$ &
 $10.6\times10^{-10}$  & 78217/339 \nl
PL              & 2.615 & 1.063 & $23.3\times10^{-10}$ & 
 ...  & ... & ... & 7545/339 \nl
PL + BB         & 2.787 & 1.073 & $19.8\times10^{-10}$ &
 1.540 & $3.11\times10^{-3}$ & $2.58\times10^{-10}$ &  1491/318 \nl
 
 
~error from $n_H$    & 0.017 & 0.043 & $0.95\times10^{-10}$ & 
 0.013 & $0.05\times10^{-3}$ & $0.06\times10^{-10}$ & ... \nl
~error from response & 0.234 & 0.446 & $5.42\times10^{-10}$ & 
 0.171 & $0.28\times10^{-3}$ & $0.22\times10^{-10}$ & ... \nl
 
\tablecomments{Fits to the 2864 second interval beginning UTC
8/7/96 01:48:19. The equivalent hydrogen column density has been
fixed at $n_H=0.18\times10^{-22}$ cm$^{-2}$. The bottom two rows show the
errors  introduced in the PL+BB fit by 1) the different values of 
$n_H$ and 2) the different response  matrices used in the fit.}
 
 
\tablenotetext{a}{Flux: erg cm$^{-2}$ s$^{-1}$ (1--20 keV).}
 
\enddata \label{table:spec_fits}
 
\end{deluxetable}


\begin{figure*}
\figurenum{1}
\epsscale{2.0}
\plotone{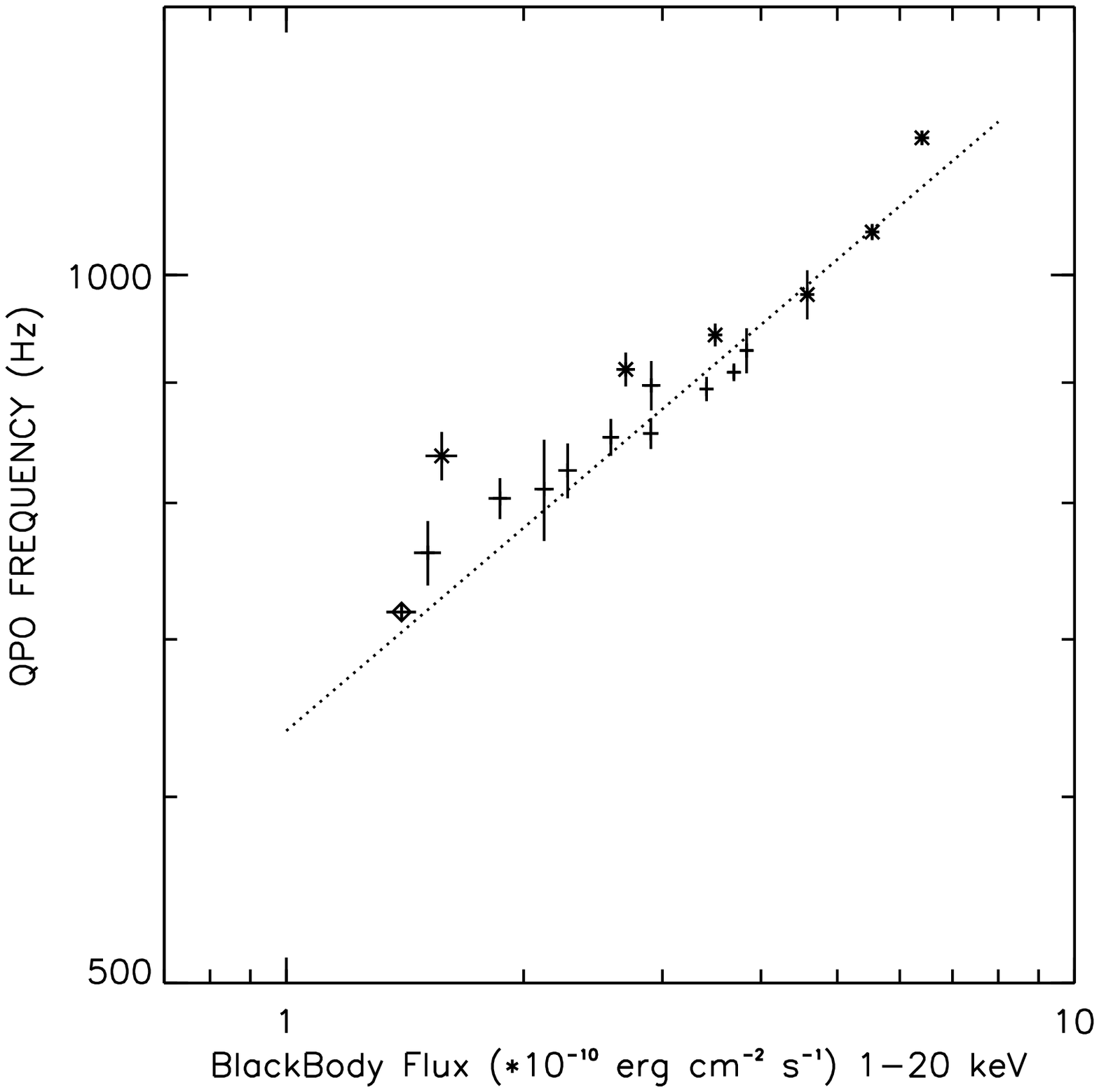} 
\caption{QPO frequency versus
flux of the blackbody spectral  component. The plusses are April
data, and the asterisks are August data. The power law fit
has a slope of $\alpha=0.29\pm0.01$. The fluxes are from a spectral
fit using a PL+BB function (Table~1), with equivalent  hydrogen
column density of $0.18\times10^{-22}$ cm$^{-2}$. The errors
plotted include all systematics except the variations due to
changing the detector response matrices. The data point at
719 Hz (diamond) is from Mendez et al. (1997) and is not used in 
determining the power law fit.}
\label{fig:freqfbb}
\end{figure*}

\begin{figure*}
\figurenum{2}
\epsscale{2.0}
\plotone{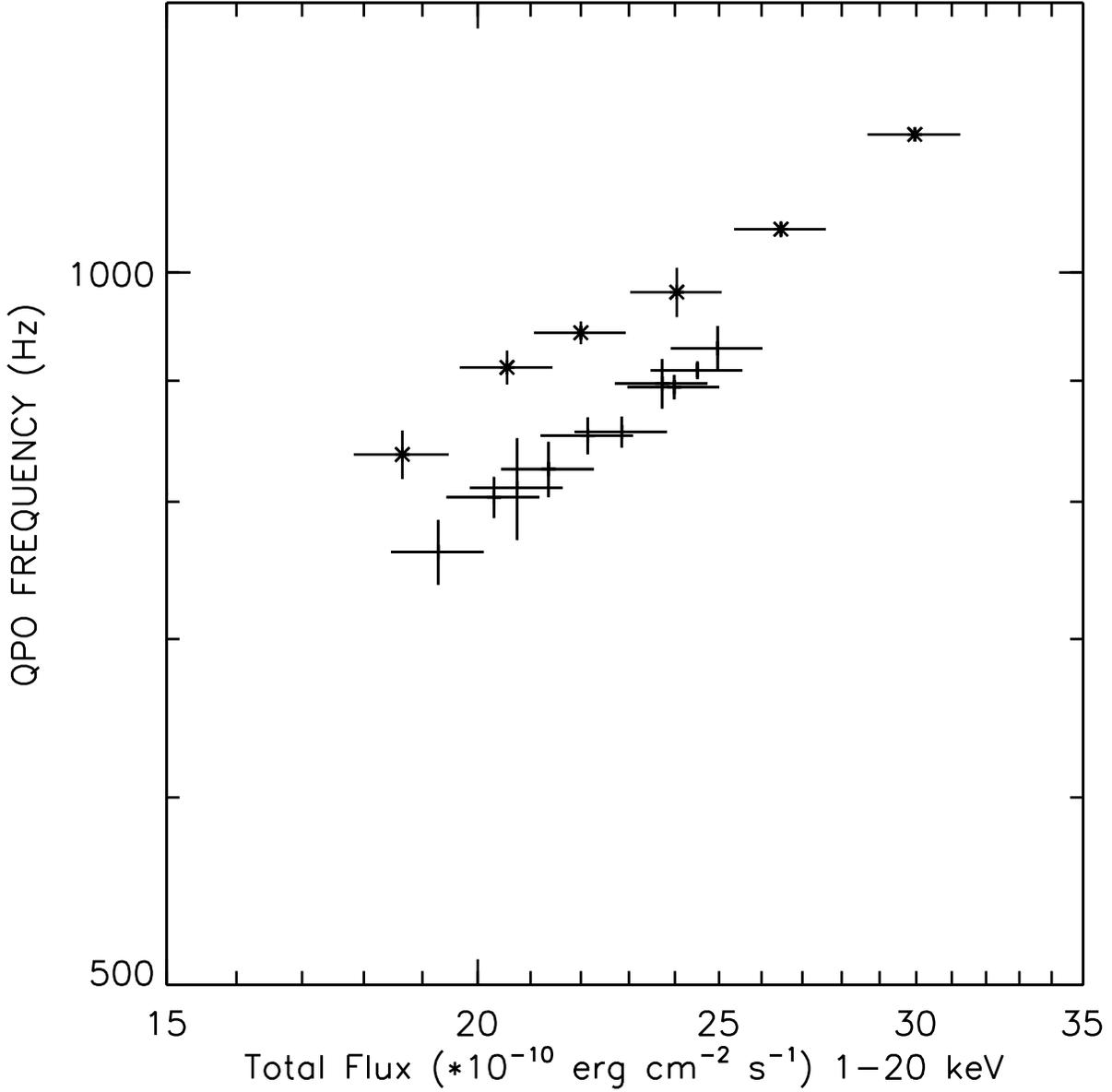}
\caption{QPO frequency versus the total energy flux, based on the 
same fits as Figure~1. The plusses are April data, and the asterisks 
are August data. The April data can be fit by a power law with 
slope, $\alpha=1.4\pm0.3$, while, for the August data, $\alpha=1.6\pm0.2$.}
\label{fig:freqftot}
\end{figure*}

\begin{figure*}
\figurenum{3}
\epsscale{2.0}
\plotone{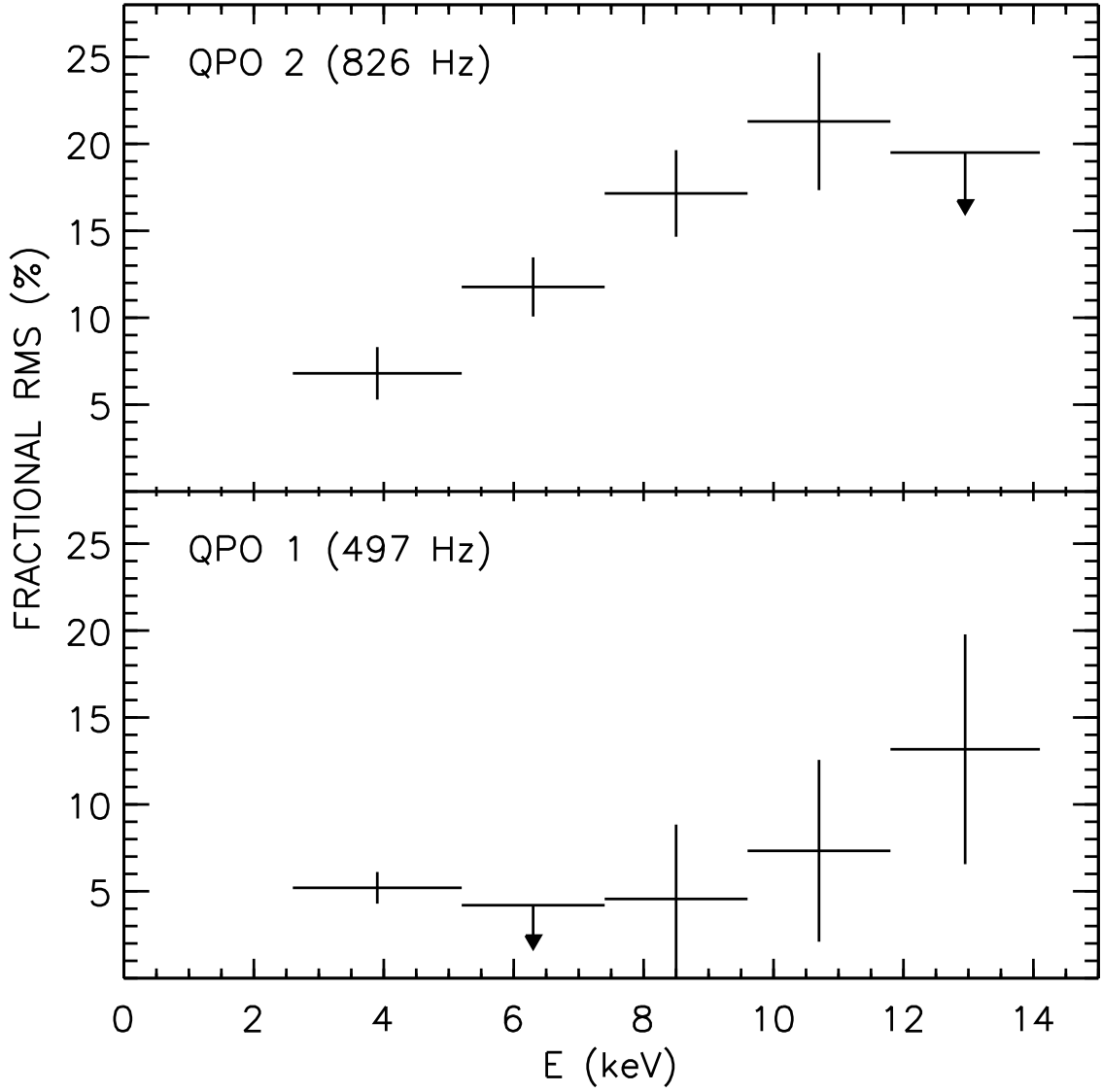}
\caption{RMS amplitude vs. energy for the
two QPOs detected in the  interval beginning UTC 4/25/96
03:11:26. The top panel shows data for  the higher frequency
QPO, while the bottom panel is for the lower frequency QPO.
Upper limits are $2\sigma$.}
\label{fig:rmsE}
\end{figure*}

\end{document}